\documentclass{moriond}

\def\Journal#1#2#3#4{{#1} {\bf #2}, #3 (#4)}

\def\NPA{{\em Nucl. Phys.} A}
\def\PLB{{\em Phys. Lett.}  B}
\def\PRL{\em Phys. Rev. Lett.}
\def\PRD{{\em Phys. Rev.} D}

\def\PRC{{\em Phys. Rev.} C}

\def\be{\begin{equation}}
\def\ee{\end{equation}}
\def\bea{\begin{eqnarray}}
\def\eea{\end{eqnarray}}

\newcommand{\Photo}{}
\begin{document}

\vspace*{4cm}
\title{MULTIPLICITY AND UNDERLYING EVENT IN ALICE: \\ AS MEASUREMENTS AND AS TOOLS TO PROBE QCD}

\author{V. ZACCOLO, ON BEHALF OF THE ALICE COLLABORATION }

\address{ INFN - Sezione di Torino \\
Via Pietro Giuria 1, 10125 Torino, Italy}

\maketitle\abstracts{ With the high collision energies at the LHC, the contributions to particle production from hard-QCD processes increase, but it remains dominated by soft-QCD processes.
Such processes challenge the theoretical models, since they are described by non-perturbative phenomenology. 
A selection of the most recent ALICE measurements of charged-particle multiplicities and the Underlying Event will be presented, focusing on model comparisons.
A summary of the current understanding of soft-QCD processes will be discussed, evaluating possible ways to further constrain theory.}
\section{Introduction}
\label{sec:intro} 
For the majority of the processes observed at the LHC non-perturbative aspects are involved.
Of specific interest are Multiple Parton Interactions (MPI) that refer to the presence of more than one hard collision, with high transverse momentum $p_{\text{T}}$.
In the following, measurements of charged-particle multiplicities and the Underlying Event (UE), constituted by semi-hard and soft events, will be presented and discussed, highlighting model comparisons.
\mbox{ALICE} is constituted by 18 different detector systems and has good momentum resolution and excellent particle identification.
The experiment is described elsewhere~\cite{alice}.
\section{Underlying Event}
\label{sec:UE}
\begin{figure}[h]
   \begin{minipage}{0.48\textwidth}
   \centering
        \includegraphics[width=\textwidth]{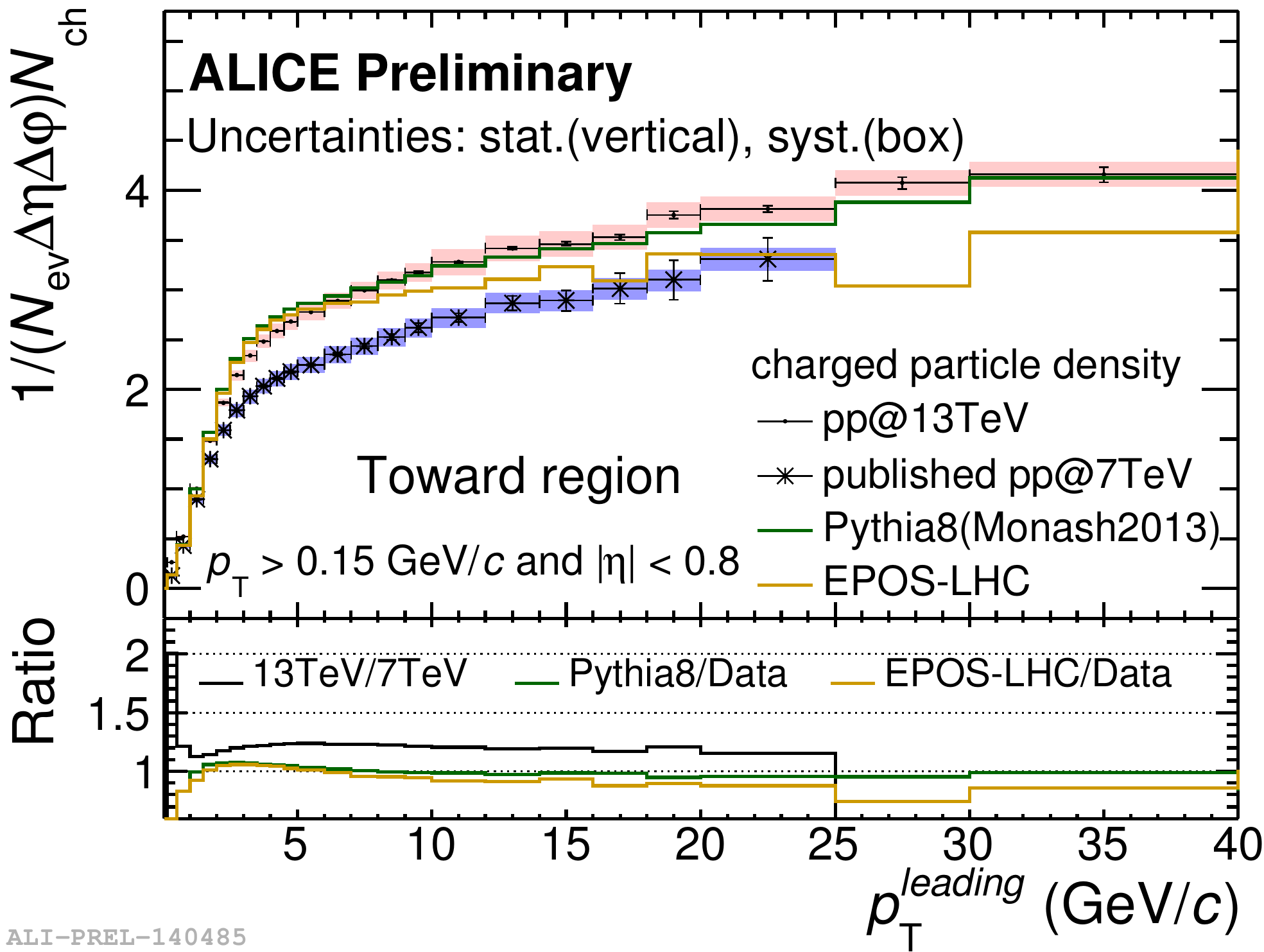}
   \end{minipage}
  \begin{minipage}{0.48\textwidth}
  \centering
       \hspace{0.1cm}
        \includegraphics[width=\textwidth]{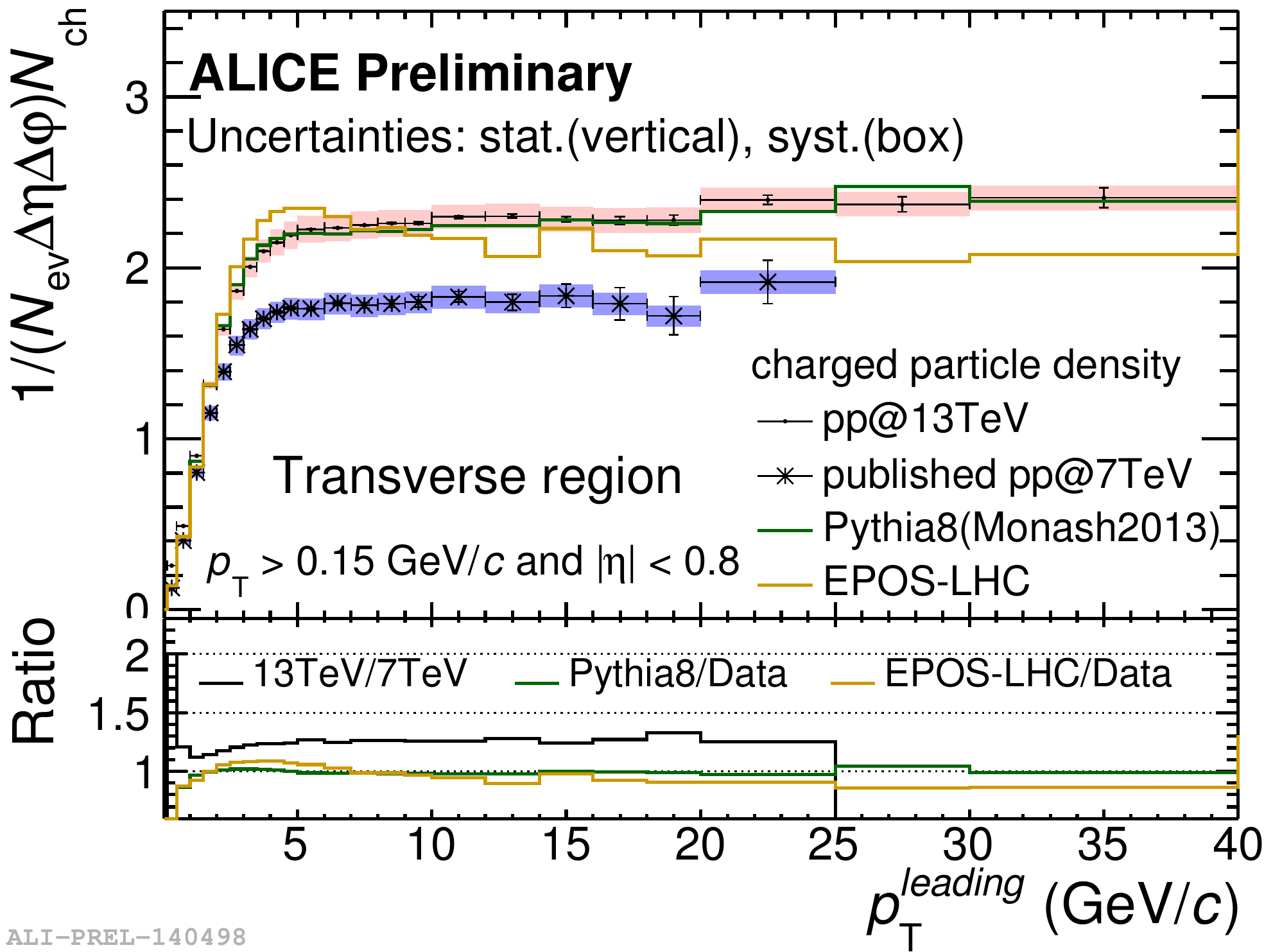}
	\end{minipage}
	\vspace{-0.4cm}
\caption{Left: Number density in the toward region in pp collisions at $\sqrt{s}=$ 13 TeV. Right: Number density in the transverse region.}\label{fig:2}	
\end{figure}

The measurement of the UE observables is crucial to separate soft and hard processes as a function of the leading track, i.e. the track with highest transverse momentum.
Measurements from ALICE exist for pp collisions at \mbox{$\sqrt{s}$ = 0.9, 7 TeV} and 13 TeV~\cite{ALICE:2011ac}.
Figure~\ref{fig:2} shows the results for the average charged-particle density as a function of the $p_{\text{T}}$ of the leading track for toward (left) and transverse (right) regions.
For the toward and away regions with respect to the leading track, where the fragmentation products from hard scattering are accumulated, the average particle density increases monotonically.
The UE is probed in the transverse region, in which the particle density grows up to a few GeV and then flattens forming a plateau.
This flattening can be attributed to the insignificance of the hard processes to the particle density at high leading-track $p_{\text{T}}$, while at low leading-track $p_{\text{T}}$ the particle density is influenced by hard processes (and eventually by MPI).
\section{Particle multiplicities}
\label{sec:mult}
\begin{figure}[h]
    \begin{minipage}{0.49\textwidth}
    \centering
        \hspace{0.4cm}
        \includegraphics[width=\textwidth]{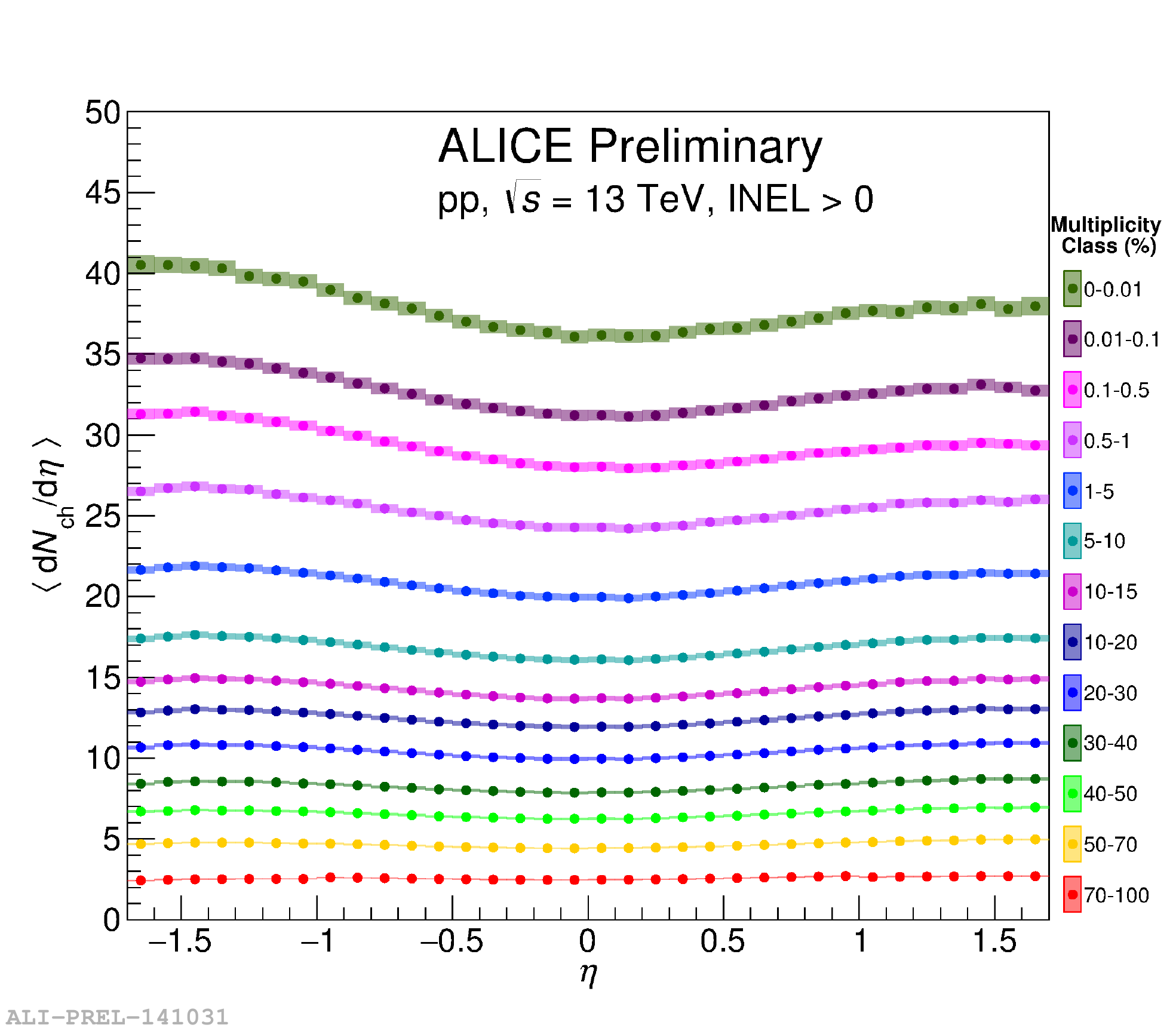}
    \end{minipage}
 \begin{minipage}{0.47\textwidth}
 \centering
        \includegraphics[width=\textwidth]{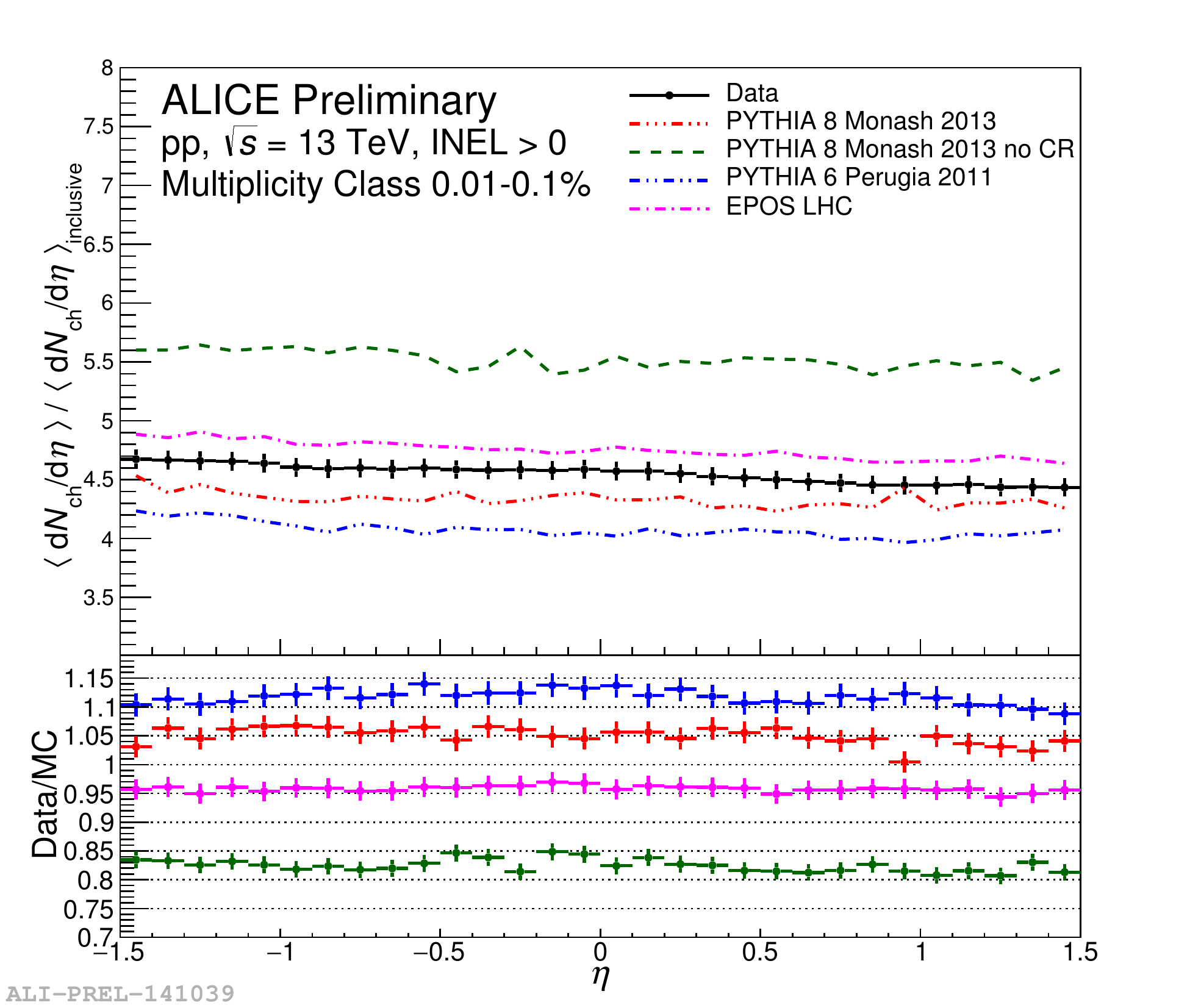}
	\end{minipage}
	\vspace{-0.4cm}
   \begin{minipage}{0.48\textwidth}
        \centering
        \includegraphics[width=\textwidth]{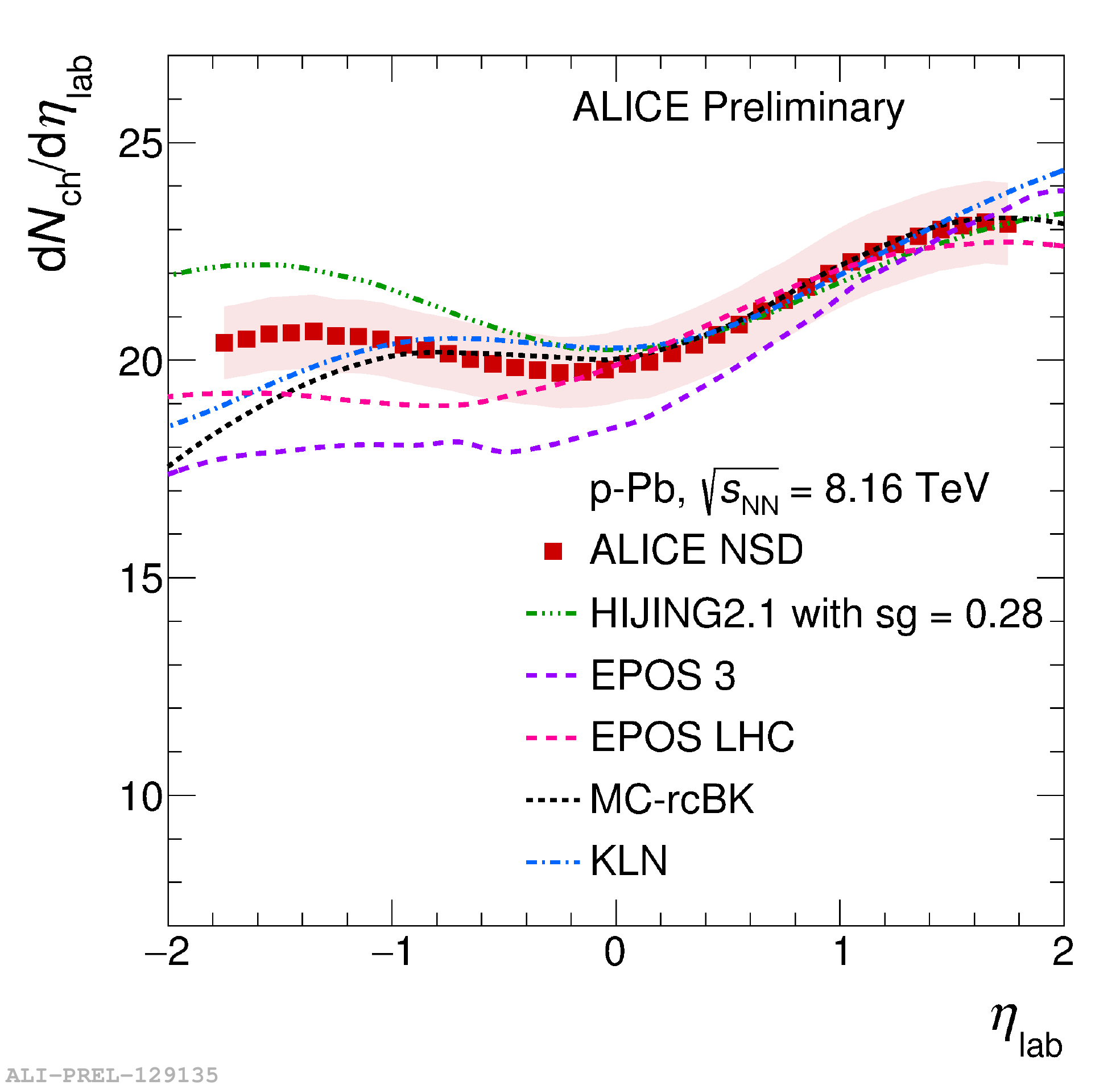}
    \end{minipage}
  \begin{minipage}{0.48\textwidth}
  \centering
        \includegraphics[width=\textwidth]{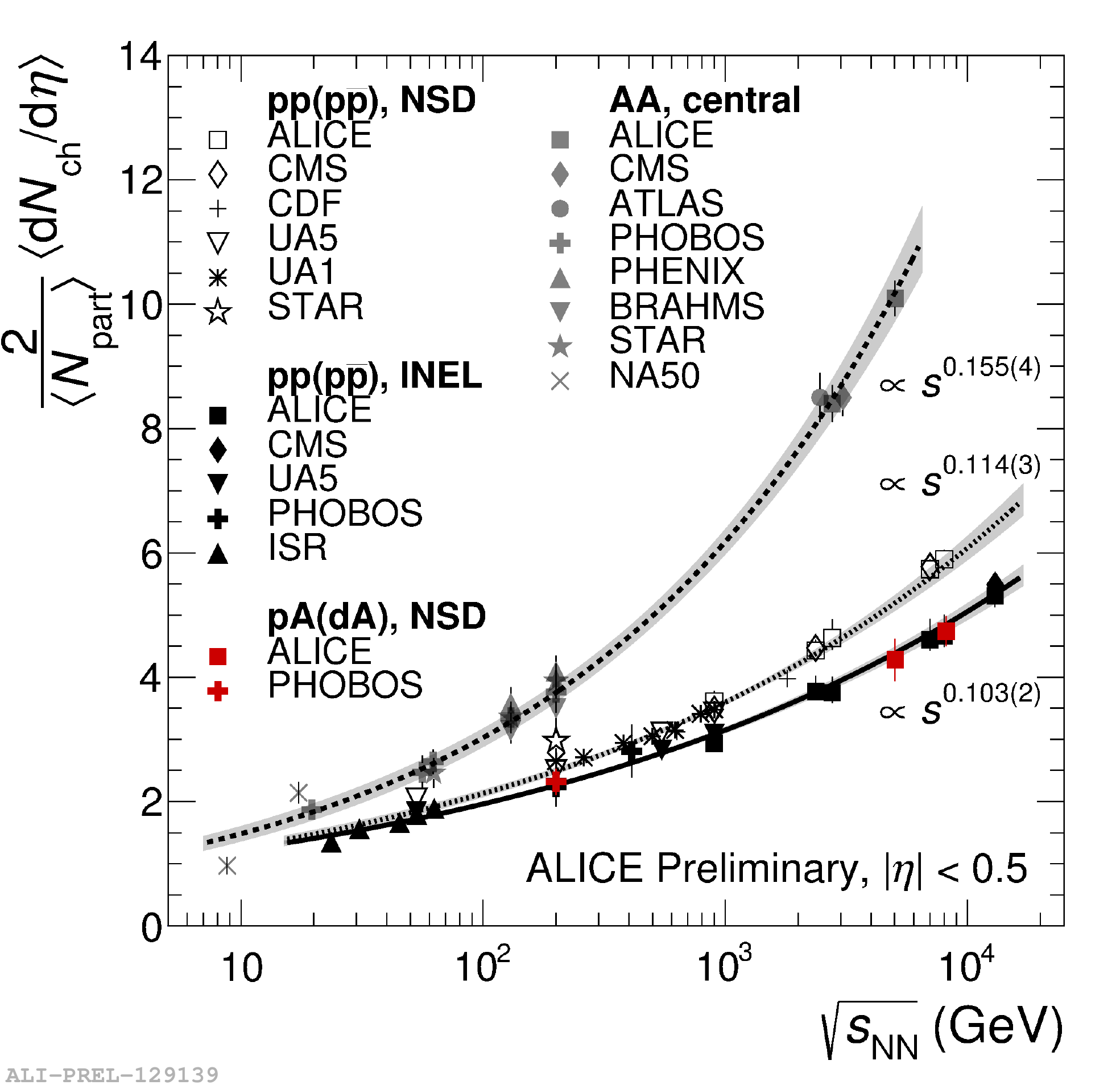}
	\end{minipage}
\caption{Top left: Pseudorapidity density of charged particles measured for pp collisions at $\sqrt{s}=$ 13 TeV in forward multiplicity slices. Top right: Monte Carlo comparisons to relative pseudorapidity density. Bottom left: $\text{d}N_{\text{ch}} / \text{d}\eta_{\rm lab}$  in p--Pb collisions at $\sqrt{s_{\text{NN}}}$=8.16 TeV. Bottom right: $\text{d}N_{\text{ch}} / \text{d}\eta$ at midrapidity as a function of $\sqrt{s_{\text{NN}}}$.}\label{fig:1}	
\end{figure}

Particle multiplicities are essential as a reference for other measurements and for tuning theoretical models.
Both the pseudorapidity density $\text{d}N_{\text{ch}} / \text{d}\eta$ and the probability $\text{P}(N_{\text{ch}})$ of charged particles have been measured by ALICE in proton--proton, pp, collisions at $\sqrt{s}$ = 0.9 to 8 TeV~\cite{Aamodt:2009aa} and at 13 TeV~\cite{Adam:2015pza}.
Figure~\ref{fig:1} top left shows the $\text{d}N_{\text{ch}} / \text{d}\eta$ as a function of the pseudorapidity in multiplicity slices derived from the V0 detector amplitude for high-multiplicity triggered data.
On the top right plot, a comparison with Monte Carlo models is performed, PYTHIA 8~\cite{Sjostrand:2007gs} with Colour Reconnection, PYTHIA 6 Perugia 2011~\cite{Skands:2010ak} and EPOS LHC~\cite{Pierog:2013ria} agree well with the data.
On the bottom right panel of Fig.~\ref{fig:1}, the $\text{d}N_{\text{ch}} / \text{d}\eta$ distribution as a function of the pseudorapidity $\eta$ in the laboratory system is shown for proton--lead, p--Pb, collisions at $\sqrt{s_{\text{NN}}}$ = 8.16 TeV. 
The number of charged particles is higher in the Pb-going side, at positive $\eta$. 
In general, models show a good agreement in the Pb-fragmentation side \cite{Deng:2010mv, Werner:2013tya, Pierog:2013ria}. 
In the p-going side, theoretical calculations that assume gluon saturation, MC-rkBK~\cite{Albacete:2012xq} and KLN~\cite{Dumitru:2011wq}, reproduce the distribution better.
On the bottom left, $\text{d}N_{\text{ch}} / \text{d}\eta$ at midrapidity is scaled by half the average number of participants calculated with a Glauber model as a function of $\sqrt{s_{\text{NN}}}$. Since the contribution from diffractive processes is negligible, the pA points agree with the pp inelastic event class. 
The rise of AA points is much steeper with respect to pp and pA. 
ALICE has measured the pseudorapidity density also for lead--lead~\cite{Adam:2015ptt} and xenon--xenon collisions~\cite{Acharya:2018hhy}, probing different system sizes and collision species, showing that the centrality-dependence distribution for the two different systems agrees up to the 10\% most central collisions.
\section{Multiplicity-dependence studies}
\label{sec:multdep}
\begin{figure}[h]
   \begin{minipage}{0.37\textwidth}
   \centering
        \includegraphics[width=\textwidth]{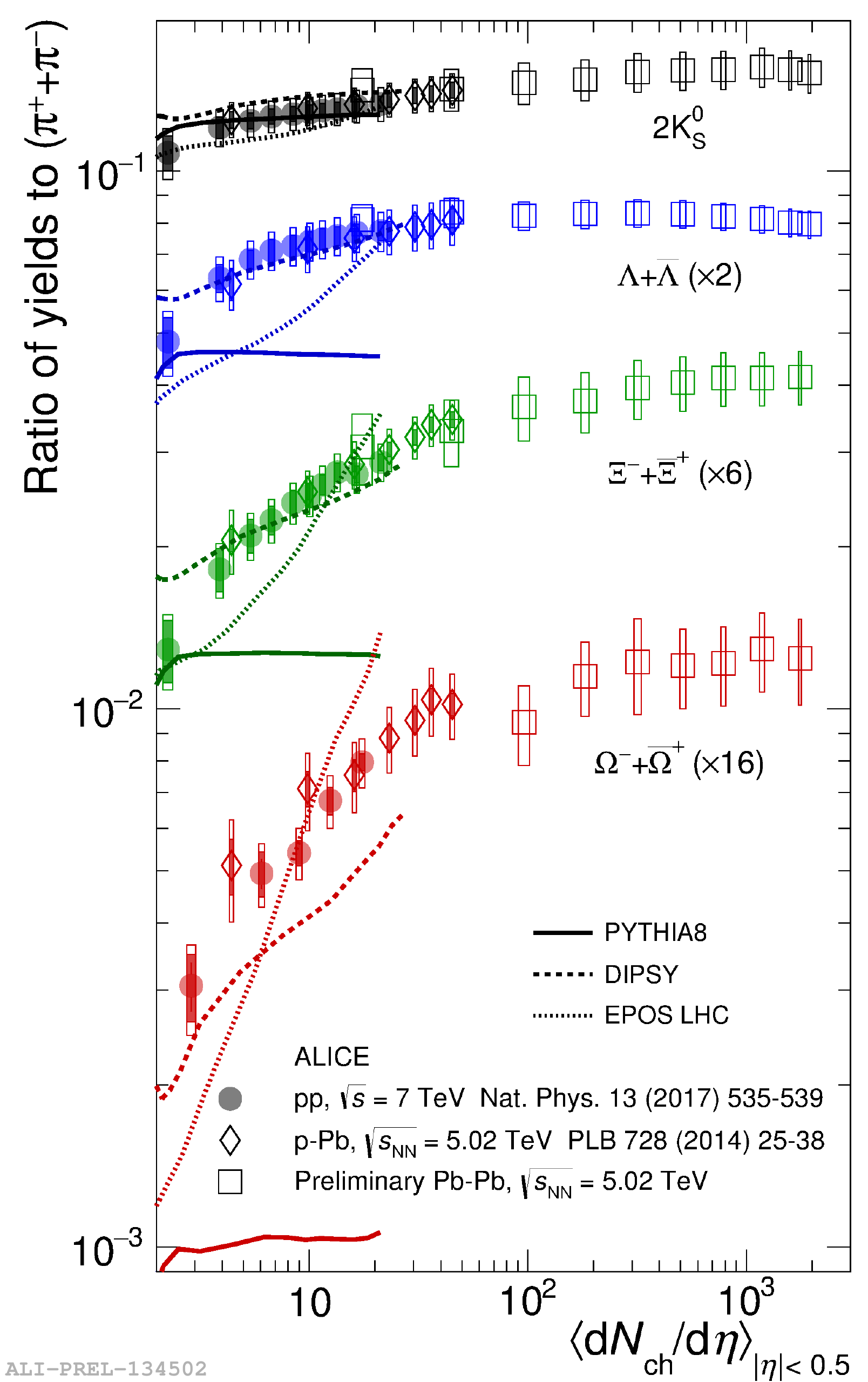}
   \end{minipage}
  \begin{minipage}{0.57\textwidth}
       \hspace{0.1cm}
       \centering
        \includegraphics[width=\textwidth]{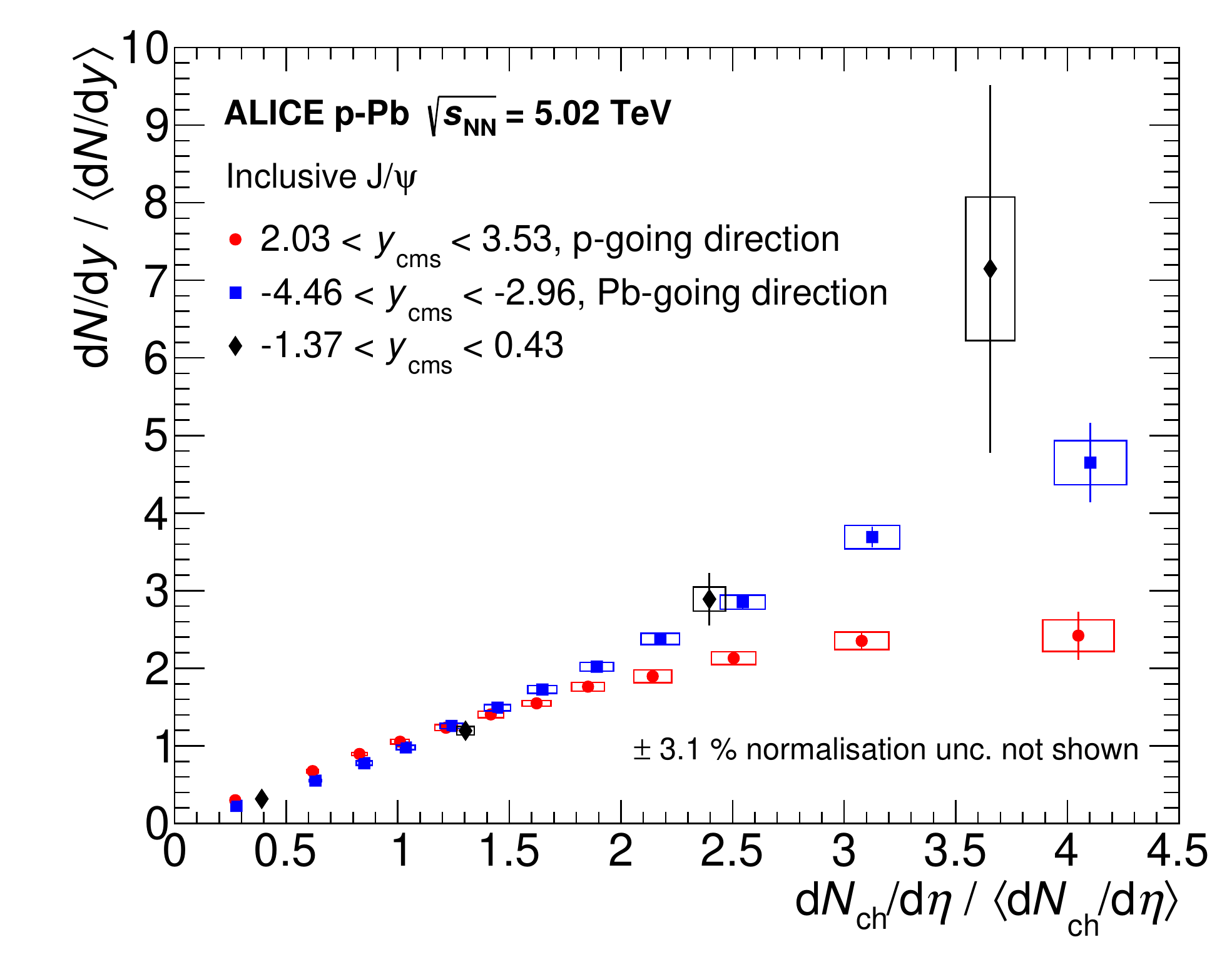}
	\end{minipage}
	\vspace{-0.4cm}
\caption{Left: Momentum-integrated yield ratios to pions as a function of multiplicity for pp~\protect\cite{ALICE:2017jyt}, p--Pb~\protect\cite{Abelev:2013haa} and Pb--Pb collisions. Right: Relative yield of inclusive J$/\Psi$ mesons as a function of relative multiplicity~\protect\cite{Adamova:2017uhu}.} \label{fig:3}	
\end{figure}

Strangeness enhancement has been used as an observable to test the formation of the Quark--Gluon Plasma in heavy-ion collisions~\cite{Koch:1986ud}. 
Nevertheless, ALICE has observed an enhancement also in high-multiplicity pp collisions when measuring the yields of strange particles~\cite{ALICE:2017jyt}.
While theoretical models are successfully describing particle multiplicity and the UE, they fail in the description of the multiplicity dependence of strange hadrons, as can be seen in Fig.~\ref{fig:3} left.
The DIPSY model~\cite{Bierlich:2015rha}, which contains the colour ropes formalism, reproduces better the data.
In Fig.~\ref{fig:3} right, instead, the J$/\Psi$ yields are presented as a function of multiplicity for p--Pb collisions~\cite{Adamova:2017uhu}.
One can observe that the yields grow faster than the diagonal for the midrapidity region (hint of multiplicity and MPI saturation), while for the forward rapidity region, where the interaction is softer, there is a hint of saturation in the J$/\Psi$ relative yield.
\section{Summary}
The charged-particle multiplicities and UE observables are described by models up to 10-20\%. 
This is a good achievement given the complexity of non-perturbative soft-QCD description.
Several measurements, like the UE and the J$/\Psi$ meson yields as a function of the multiplicity, hint to saturation of MPI at high multiplicity and high $p_{\text{T}}$.
Progress has been made in the description of the multiplicity dependence of strange hadron production, but the models are still challenged.
The multiplicity and UE measurements have significantly improved our phenomenological understanding of high-energy collisions. 
Nevertheless, further constraints can still be posed, e.g. probing QCD using the UE to test non-perturbative dynamics excluding the hard sector.

\section*{References}
\begin{thebibliography}{99}
\bibitem{alice}K. Aamodt {\it et al.} [ALICE Collaboration], \Journal{\em JINST}{3}{S08002}{2008}. 
\bibitem{ALICE:2011ac}B. Abelev {\it et al.} [ALICE Collaboration], \Journal{\em JHEP}{1207}{116}{2012}.
\bibitem{Aamodt:2009aa}K. Aamodt {\it et al.} [ALICE Collaboration], \Journal{{\em Eur. Phys. J.} C}{65}{111}{2010}; K. Aamodt {\it et al.} [ALICE Collaboration], \Journal{{\em Eur. Phys. J.} C}{68}{89}{2010}; K. Aamodt {\it et al.} [ALICE Collaboration], \Journal{{\em Eur. Phys. J.} C}{68}{345}{2010}; J. Adam{\it et al.} [ALICE Collaboration], \Journal{{\em Eur. Phys. J.} C}{77}{no.1, 33}{2017}; S. Acharya {\it et al.} [ALICE Collaboration], \Journal{{\em Eur. Phys. J.} C}{77}{no.12, 852}{2017}.
\bibitem{Adam:2015pza}J. Adam{\it et al.} [ALICE Collaboration], \Journal{\PLB}{753}{319}{2016}.
\bibitem{Sjostrand:2007gs}T.~Sjostrand, S.~Mrenna and P.~Z.~Skands, \Journal{\em Comput. Phys. Commun.}{178}{852}{2008}.
\bibitem{Skands:2010ak}P.~Z.~Skands, \Journal{\PRD}{82}{074018}{2010}.
\bibitem{Pierog:2013ria}T.~Pierog, I.~Karpenko, J.~M.~Katzy, E.~Yatsenko and K.~Werner, \Journal{\PRC}{92}{no.3, 034906}{2015}.
\bibitem{Deng:2010mv}W.~T.~Deng, X.~N.~Wang and R.~Xu,  \Journal{\PRC}{83}{014915}{2011}.
\bibitem{Werner:2013tya}K.~Werner, B.~Guiot, I.~Karpenko and T.~Pierog,  \Journal{\PRC}{93}{no.6, 064903}{2014}.
\bibitem{Albacete:2012xq}J.~L.~Albacete, A.~Dumitru, H.~Fujii and Y.~Nara, \Journal{\NPA}{897}{1}{2013}.
\bibitem{Dumitru:2011wq}A.~Dumitru, D.~E.~Kharzeev, E.~M.~Levin and Y.~Nara, \Journal{\PRC}{85}{044920}{2012}.
\bibitem{Adam:2015ptt}J. Adam {\it et al.} [ALICE Collaboration], \Journal{\PRL}{116}{no.22,  222302}{2016}; K. Aamodt {\it et al.} [ALICE Collaboration], \Journal{\PRL}{106}{ 032301}{2011}; K. Aamodt {\it et al.} [ALICE Collaboration], \Journal{\PRL}{105}{252301}{2010}.
\bibitem{Acharya:2018hhy}S. Acharya {\it et al.} [ALICE Collaboration], \texttt{arXiv:1805.04432 [nucl-ex]}.
\bibitem{Koch:1986ud}P. Koch, B. Muller, and J. Rafelski, \Journal{\em Phys. Rept}{142}{167-262}{1986}.
\bibitem{ALICE:2017jyt}J. Adam{\it et al.} [ALICE Collaboration], \Journal{\em Nature Phys.}{13}{535-539}{2017}.
\bibitem{Abelev:2013haa}B. Abelev {\it et al.} [ALICE Collaboration], \Journal{\PLB}{728}{25-38}{2014}.
\bibitem{Adamova:2017uhu}D. Adamova {\it et al.} [ALICE Collaboration], \Journal{\PLB}{776}{91-104}{2018}.
\bibitem{Bierlich:2015rha}C. Bierlich and J.R. Christiansen, \Journal{\PRD}{93}{no.9, 094010}{2015}.
\end{thebibliography}
\end{document}